\documentclass[aps,pra,superscriptaddress,notitlepage,twocolumn]{revtex4-2}

\usepackage{mathrsfs}
\usepackage{amsfonts}
\usepackage{amssymb}
\usepackage{amsmath}
\usepackage{bm}
\usepackage{graphicx}
\usepackage{sidecap}
\usepackage{color}
\usepackage[colorlinks=true,citecolor=blue,linkcolor=cyan]{hyperref}
\usepackage{subeqnarray}
\usepackage{verbatim}
\usepackage{ulem}
\usepackage{cases}
\usepackage{cancel}
\usepackage{framed}
\usepackage{tikz}
\usepackage{multirow}

\newcommand{\ket}[1]{\vert #1 \rangle}
\newcommand{\bra}[1]{\langle #1 \vert}
\newcommand{\ketbra}[2]{\vert #1 \rangle \langle #2 \vert}
\newcommand{\braket}[2]{\langle #1 \vert #2 \rangle}

\begin{document}
\title{Speeding up adiabatic holonomic quantum gates via $\pi$-pulse modulation}
\author{Jiang Zhang}

\affiliation{Beijing Academy of Quantum Information Sciences, Beijing 100193, China}
\author{Tonghao Xing}

\email{xingth@baqis.ac.cn}
\affiliation{Beijing Academy of Quantum Information Sciences, Beijing 100193, China}
\affiliation{State Key Laboratory of Low-Dimensional Quantum Physics and Department of Physics, Tsinghua University, Beijing 100084, China}
\author{Guilu Long}
\email{gllong@mail.tsinghua.edu.cn}
\affiliation{Beijing Academy of Quantum Information Sciences, Beijing 100193, China}
\affiliation{State Key Laboratory of Low-Dimensional Quantum Physics and Department of Physics, Tsinghua University, Beijing 100084, China}
\affiliation{Frontier Science Center for Quantum Information, Beijing 100084, China}
\affiliation{Beijing National Research Center for Information Science and Technology, Beijing 100084, China}

\begin{abstract}
Holonomic quantum computation (HQC) offers an inherently robust approach to quantum gate implementation by exploiting quantum holonomies. While adiabatic HQC benefits from robustness against certain control errors, its long runtime limits practical utility due to increased exposure to environmental noise. Nonadiabatic HQC addresses this issue by enabling faster gate operations but compromises robustness. In this work, we propose a scheme for fast holonomic quantum gates based on the $\pi$-pulse method, which accelerates adiabatic evolution while preserving its robustness. By guiding the system Hamiltonian along geodesic paths in the parameter space and applying phase-modulating $\pi$ pulses at discrete points, we realize a universal set of holonomic gates beyond the conventional adiabatic limit. Our scheme allows for arbitrary single-qubit and two-qubit controlled gates within a single-loop evolution and provides additional tunable parameters for flexible gate design. These results demonstrate a promising path toward high-fidelity, fast, and robust quantum computation.

\end{abstract}
\maketitle

\section{Introduction and motivation}

Holonomic quantum computation (HQC) is the concept of using quantum holonomy to construct robust quantum gates for universal quantum computation (for a review, see \cite{zhang2023geometric,hao2021quantum}; for  related work on geometric quantum computation, see \cite{zhang2023geometric,jones2000geometric,wang2001nonadiabatic,zhu02}). Originally, HQC was proposed based on the adiabatic evolution of an energy-degenerate subspace, where the corresponding dynamical evolution is trivial, while the adiabatic quantum holonomy—i.e., the Wilczek-Zee phase \cite{wilczek1984appearance}—acts on the subspace to realize quantum gates \cite{zanardi1999holonomic}.
A concrete set of universal adiabatic holonomic gates was proposed for trapped ions confined in a linear Paul trap \cite{duan2001geometric}.
The tripod scheme introduced in this pioneering work has been extended to various physical systems, such as superconducting systems \cite{faoro2003non} and semiconductor systems \cite{solinas2003semiconductor}, and has been experimentally demonstrated with trapped ions \cite{toyoda2013realization}.

Adiabatic evolution is inherently robust against control errors \cite{childs2001robustness}. However, realizing adiabatic HQC typically requires a long runtime to satisfy the adiabatic condition \cite{messiah2014quantum}. This prolonged evolution exposes the system excessively to environmental errors, which can distort the quantum evolution.
To circumvent this issue, nonadiabatic HQC \cite{sjoqvist2012non,xu2012non} was proposed based on nonadiabatic quantum holonomies \cite{anadan1988non} in a three-level $\Lambda$-model configuration. Following this approach, several improved schemes have been developed, including the single-shot scheme \cite{xu2015non,sjoqvist2016non}, which realizes arbitrary single-qubit gates in a single loop; the path-shortening scheme \cite{xu2018path}, which reduces the total evolution time; and the multi-pulse scheme based on resonant control \cite{herterich2016single}.
Moreover, a four-level structure beyond the $\Lambda$ model has been proposed \cite{mousolou2014universal}, enabling the implementation of holonomic gates directly with qubits \cite{gurkan2015realization,zhang2018holonomic,wang2020dephasing}, thus avoiding the need for multi-level systems. Nonadiabatic HQC has since undergone significant development, both theoretically \cite{liang2014nonadiabatic,zhang2014quantum,zheng2014fault,zheng2015fault,xue2015universal,wang2016holonomic,song2016shortcuts,zhao17rydberg,zhao2017single,kang2018nonadiabatic,chen2018nonadiabatic,zhao2019nonadiabatic,ramberg2019environment,liu2019plug,zhang2019searching,wu2020holonomic,xing2020nonadiabatic,xing2020nonadiabatic2,xu2021realizing,zhao2021dynamical,xing2021realization,liang2022composite,xu2022realizing,xu2025reducing} and experimentally \cite{abdumalikov2013experimental,feng2013experimental,zu2014experimental,arroyo2014room,zhou2017holonomic,sekiguchi2017optical,li2017experimental,xu2018single,nagata2018universal,ishida2018universal,danilin2018experimental,zhu2019single,egger19,zhang2019single,ai2020experimental,ai2022experimental,li2023dynamical}.
Moreover, owing to its compatibility with various noise-resilient strategies, HQC has been integrated with other approaches such as decoherence-free subspaces (DFS) \cite{xu2012non,liang2014nonadiabatic,zhao2017single}, noiseless subsystems (NS) \cite{zhang2014quantum}, dynamical decoupling (DD) \cite{zhao2021dynamical}, and quantum error correction (QEC) \cite{zheng2014fault,zheng2015fault,zhang2018holonomic,wu2020holonomic}.
Nevertheless, in nonadiabatic holonomic implementations, the states of the system do not follow the instantaneous eigenstates of the system Hamiltonian. As a result, nonadiabatic holonomic gates generally do not inherit the same robustness properties as their adiabatic counterparts.

Another approach to realizing fast holonomic gates is to accelerate the corresponding adiabatic evolution using alternative methods. One such method is the shortcut to adiabaticity \cite{berry2009transitionless,chen2010shortcut,del2013shortcuts,zhang2015fast,santos2017generalized,guery2019shortcuts,del2019focus,abah2020quantum,chen2021shortcuts,schaff2010fast,bason2012high,deng2018superadiabatic,zhang2013experimental,kolbl2019initialization,an2016shortcuts,vepsalainen2019superadiabatic}, which introduces an additional Hamiltonian to suppress nonadiabatic excitations arising from the rapid change of the system Hamiltonian. Despite enabling fast evolution, in such schemes the instantaneous state of the system no longer remains an eigenstate of the total Hamiltonian. As a result, the accelerated adiabatic holonomic gates do not retain the intrinsic robustness of adiabatic evolution either. Moreover, the required additional Hamiltonian imposes stringent requirements on system control, thereby reducing the experimental feasibility of this approach \cite{ibanez2012multiple,martinez2014shortcuts,li2016shortcut,baksic2016speeding,du2016experimental,zhou2017accelerated}.
To combine robustness with high-speed operation in HQC, it is therefore desirable for the quantum state to follow the instantaneous eigenstate of the Hamiltonian while evolving more rapidly through an appropriate control scheme.

In this work, we propose the use of the $\pi$-pulse scheme \cite{wang2016necessary} to enable the fast implementation of adiabatic holonomic gates.
We begin by showing that cyclic adiabatic evolution in three-level systems can generate a universal set of holonomic gates.
We then demonstrate that this universal set can be realized by steering the system Hamiltonian along geodesic lines in the relevant parameter spaces.
Adiabatic evolution along these geodesic paths can be accelerated using the $\pi$-pulse scheme, which applies a control Hamiltonian at discrete points along the path.
The scheme is based on a necessary and sufficient condition for adiabaticity, ensuring that the system still follows the desired adiabatic trajectory even under fast driving \cite{wang2016necessary}.
The underlying idea is to compensate for nonadiabatic transitions through phase modulation induced by $\pi$ pulses acting on the eigenstates of the Hamiltonian.
Compared with standard nonadiabatic holonomic gates, our proposed adiabatic holonomic gates based on $\pi$ pulses retain the robustness inherent in adiabatic evolution, while providing additional controllable parameters that enhance flexibility in gate design.
For single-qubit gates, arbitrary holonomic gates can be realized within a single loop in the resonant regime. More importantly, for two-qubit gates, arbitrary controlled gates can also be achieved within a single loop—an advantage not available in nonadiabatic HQC schemes.
We explicitly demonstrate the implementation of a universal set of fast holonomic gates beyond the conventional adiabatic limit.
Our scheme opens a promising route toward high-fidelity, fast, and robust quantum gate operations for fault-tolerant quantum computation \cite{terhal2015quantum,cai2021bosonic}.

\section{Evolution operator in the adiabatic limit}
Consider a quantum system driven by a time-dependent Hamiltonian $H(t)$.
The corresponding time evolution operator $U(t)$ takes the form of
\begin{equation}
	i\dot{U}(t)=H(t)U(t).
\end{equation}
If we use a parameter $s=t/\tau$ with $\tau$ being the total evolution time to describe the changing rate of $H(t)$, the equation of $U(t)$ can be reformed as
\begin{equation}
	i\dot{U}(s)=\tau H(s)U(s)
\end{equation}
In this setting, the Hamiltonian $H(s)$ can be shown as
\begin{equation}
	H(s)= \sum _{n}E_{n}(s)| \varphi _{n}(s)\rangle \langle \varphi _{n}(s)|,
\end{equation}
where $E_n(s)$ are the eigenvalues and $\ket{\varphi_n(s)}$ are the associated eigenstates.

The adiabatic theorem states that if the system is initially in one of the eigenstates $\ket{\varphi_n(0)}$ and the Hamiltonian $H(s)$ changes adiabatically, the instantaneous state of the system will follow $\ket{\varphi_n(t)}$.
However, as the system evolves, a phase will accumulate for $\ket{\varphi_n(t)}$.
To describe the adiabatic evolution explicitly, we define $\ket{\tilde{\varphi}_n(s)}=e^{-i\tau\alpha_n(s)+i\gamma_n(s)}\ket{\varphi_n(s)}$, where $-\alpha_n(s)=-\int_{0}^{s}E_n(s')\mathrm{d}s$ is the dynamical phase and $\gamma_n(s)=i\int_{0}^{s}\langle\varphi_n(s')|\dot{\varphi}_n(s)\rangle \mathrm{d}s'$ is the geometric phase. Then the adiabatic evolution operator $U_A(s)$ can be shown as
\begin{equation}
	U_A(s)= \sum _{n}| \tilde{\varphi}_{n}(s)\rangle \langle \tilde{\varphi}_{n}(0)|= \sum _{n}| \tilde{\varphi}_{n}(s)\rangle \langle \varphi _{n}(0)|.
\end{equation}

In the following, we are going to prove that, when $\tau$ goes to infinity, $U_A(s)$ is the evolution operator generated by $H(s)$, i.e.,
\begin{equation}
	\lim_{\tau \rightarrow +\infty}U_A^{\dagger}(s)U(s)=I,
\end{equation}
where $I$ is the identity acting on the system.
To this end, we define the transition operator $U_T(s) = U_A^{\dagger}(s)U(s)$ to characterize the difference between the real evolution operator $U(s)$ and adiabatic evolution operator $U_A(s)$. Accordingly, we have
\begin{equation}\label{dw}
	i \dot{U}_T(s)=[\tau U_A^{\dagger}(s)H(s)U_A(s)+i\dot{U}_A^{\dagger}(s)U_A(s)]U_T(s),
\end{equation}
where we adopt $\dot{K}\equiv\mathrm{d}K/\mathrm{d}s$ with $K$ being a matrix or vector here and hereafter.
By setting $H_T=\tau U_A^{\dagger}HU_A+i\dot{U}_A^{\dagger}U_A$, we obtain that
\begin{align}\label{k}
	H_T(s)=& \sum _ { k \neq j } i e ^ { i \tau[ \alpha _ { k } ( s ) - \alpha _ { j } ( s ) ] + i [ \gamma _ { j } ( s ) - \gamma _ { k } ( s ) ] }  \nonumber \\
	& \cdot\langle \dot{\varphi}_{k}(s)| \varphi _{j}(s)\rangle| \varphi _{k}(0)\rangle \langle \varphi _{j}(0)|
\end{align}
In the basis of $\{\ket{\varphi_n(0)}\}$, $H_T(s)$ comprises only off-diagonal elements.

For a general $H(s)$, $U_T(s)$ can be expanded using Dyson series as
\begin{align}
	U_T(s)=&1-i\int_{0}^{s}\mathrm{d}s_1H_T(s_1)\notag\\
	&-\int_0^s\mathrm{d}s_1\int_0^{s_1}\mathrm{d}s_2H_T(s_1)H_T(s_2)+\cdots
\end{align}
where we have only listed the first two non-trivial terms and $s_1>s_2$.

By defining $F(s)=\int_{0}^{s}\mathrm{d}s_1H_T(s_1)$ and using Hilbert-Schmidt norm, we have
\begin{align}\label{w1}
	||U_T(s)\!-\!1|| \leq &||F(s)||\!+\!\int _{0}^{s}||H_T(s_1)||\cdot||F(s_1)||\mathrm{d}s_1\!+\!\cdots,
\end{align}
where $||U_T(s)||=1$ since it is an unitary operator. It is clear that if $\lim_{\tau\rightarrow +\infty}||F(s)||=0$, the norm of the higher-order terms in $U_T(s)$ are all higher-order infinitesimals.
Therefore, $\lim_{\tau\rightarrow +\infty}||U_T(s)-1||=0$, i.e., $\lim_{\tau\rightarrow +\infty}U_T(s)=1$, implying that
there is no excitation between different eigenstates. In Appendix \ref{appa}, we show in detail that the limit $\lim_{\tau\rightarrow +\infty}||F(s)||=0$ hold.

\section{Adiabatic condition for finite evolution time}
\begin{figure}
	\centering
	\includegraphics[width=0.5\textwidth]{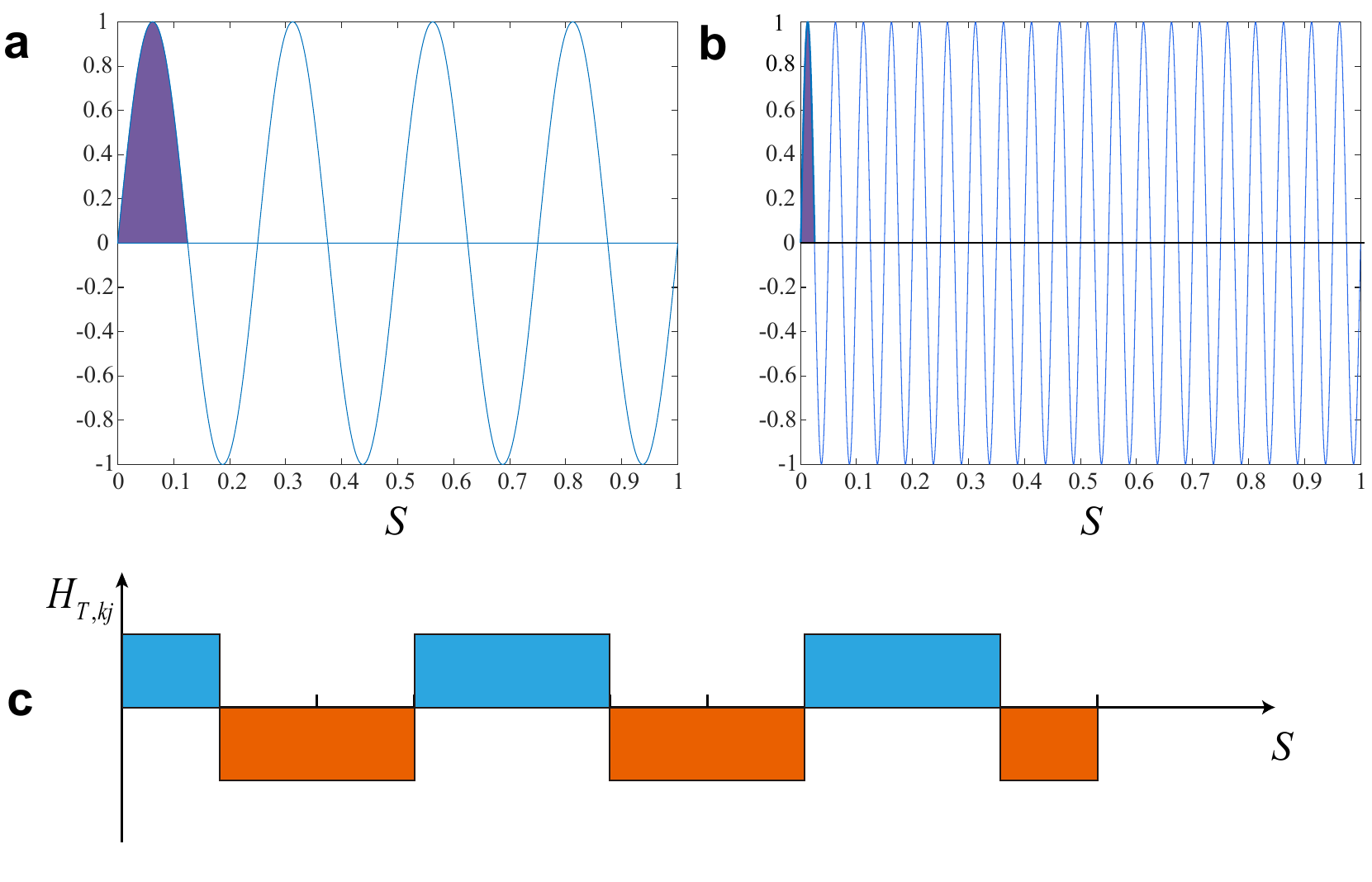}\\
	\caption{\textbf{Schematic diagram of using phase oscillations to suppress nonadiabatic transitions.} For (a) and (b), The $x$-axis represents the parameter $s$ (ranging from 0 to 1), and the $y$-axis shows the imaginary part of $e^{i\tau\Delta Es}$. We set $\tau\Delta E=8\pi$ in (a) and set $\tau\Delta E=40\pi$ in (b). The maximum value of $\|\int_0^se^{i\tau\Delta Es'}\mathrm{d}s'\|$ is about 0.08 in (a) and 0.016 in (b). (c) The $\pi$-pulse control scheme to suppress nonadiabatic transitions. The nonadiabatic transition accumulated during the first time interval (represented by the first blue rectangle) is compensated by the transition accumulated in the second time interval (represented by the first red rectangle). The residual nonadiabatic transition from the first red rectangle is subsequently canceled by the second blue rectangle, and this pattern continues. Over the entire evolution process, as long as the total area of the blue rectangles  equals that of the red rectangles, the nonadiabatic transition is completely eliminated.}\label{fig:error}
\end{figure}

In practice, the total evolution time $\tau$ cannot be too long since a quantum system is limited by its decoherence time. To design experimentally feasible quantum gates, we need to consider the adiabatic evolution that can be achieved in finite time.
One can see from Eq.~(\ref{w1}) that the transitions between different eigenstates
can be suppressed when the following condition is satisfied,
\begin{equation}
	F_{kj}(s)= \int _{0}^{s}H_{T,kj}(s^{\prime})\mathrm{d}s^{\prime}\approx 0 \quad \forall s\in[0,1],
\end{equation}
where $H_{T,kj}(s)=\bra{\varphi_k(0)}H_T(s)\ket{\varphi_j(0)}$.
To interpret the underlying physical picture behind the above condition, we rewrite $H_{T,kj}$ as
\begin{equation}
	H_{T,kj}=P_{kj}G_{kj}=ie^{i\tau \left[ \alpha _{k}(s)- \alpha _{j}(s)\right] +i \left[ \gamma _{j}(s)- \gamma _{k}(s)\right]}\langle\dot{\varphi}_{k}| \varphi _{j}\rangle ,
\end{equation}
where $P_{kj}=e^{i\tau \left[ \alpha _{k}(s)- \alpha _{j}(s)\right] +i \left[ \gamma _{j}(s)- \gamma _{k}(s)\right]}$ and $G_{kj}=i\langle\dot{\varphi}_{k}| \varphi _{j}\rangle$.
Now assuming that $G_{kj}$ is time-independent or varies  sufficient slowly compared with $P_{kj}$, the integral of $H_{T,kj}$ fluctuates as $P_{kj}$ periodically changes. Therefore, the maximum of $||U_T(s)-1||$ (the derivation of the evolution from adiabatic evolution) is determined by how fast $P_{kj}$ changes. In Fig.~\ref{fig:error}, we explicitly show that when the dynamical phase difference fluctuates faster, the maximum norm of its integral decreases linearly with how many time the phase fluctuations in $\tau$. Therefore, $U_T(s)$ better approximates the identity when $F_{kj}(s)$ oscillates more rapidly for a given $\tau$ (see Fig.~\ref{fig:error}).

The above physical picture inspires the $\pi$-pulse control scheme to realize adiabatic process in a fast way. The basic idea of $\pi$ pulse based adiabatic evolution is similar to eliminating environment noise with  dynamical decoupling: the system evolves rapidly over an  interval $\delta\tau_1$ so a nonadiabatic transition $U_T^1$ is accumulated; then a $\pi$ pulse is applied to the system between corresponding pairs of eigenstates so that $H_T(s)$ becomes $-H_T(s)$; evolve the system for another time $\delta\tau_2$ to obtain another nonadiabatic transition $U_T^2$ that cancel $U_T^1$, bringing the evolution closer to the identity, and so on (see Fig.~\ref{fig:error}(c)). The merit of the $\pi$ pulse based adiabatic evolution is that no additional control Hamiltonian is needed, and thus the instantaneous state of the system always follows the system Hamiltonian, retaining the robustness of adiabatic evolution. The main difference between dynamical decoupling and the $\pi$ pulse based on adiabatic evolution is that dynamical decoupling aims to reverse the system-environment interaction while the latter reverses the nonadiabatic transition matrix $H_T$.

\section{Using fast adiabatic evolution to build holonomic quantum gates}
When an adiabatic evolution completes a closed path in the parameter space, a corresponding holonomic gate is generated. We will show in the following that under certain conditions, this kind of holonomic gates can perform universal quantum computation based on a set of three-level quantum systems.
Compared with standard nonadiabatic holonomic gates, the proposed adiabatic holonomic gates based on $\pi$ pulses retain the robustness inherent in adiabatic evolution, while providing additional controllable parameters that enhance flexibility in gate design.

\subsection{Single-qubit holonomic gate}
\begin{figure}
	\centering
	\includegraphics[width=0.5\textwidth]{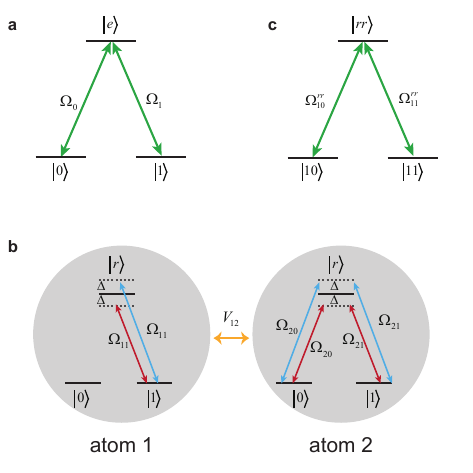}\\
	\caption{\textbf{Level structures for the single- and two-holonomic gates.} (a) The three-level $\Lambda$-model structure for realizing single-qubit holonomic gates. The two ground  states $\ket{0}$ and $\ket{1}$ are coupled to the excited state $\ket{e}$ through two pulses $\Omega_0=\Omega(t)\sin(\theta/2)e^{i\phi}$ and $\Omega_1=-\Omega(t)\cos(\theta/2)$, respectively. (b) Level structure for the two Rydberg atoms used to implement two-qubit holonomic gates. (c) Effective level structure derived from (b). In the three-level subspace, the two ground states $\ket{10}$ and $\ket{11}$ are coupled to $\ket{rr}$ via $\Omega_{10}^{rr}=2\Omega_{11}\Omega_{20}/\Delta$ and $\Omega_{11}^{rr}=2\Omega_{11}\Omega_{21}/\Delta$, respectively.}\label{fig:levels}
\end{figure}
Now assume that we can find a proper Hamiltonian $H(t)$ that satisfies the above condition so that transitions
between different eigenstates are suppressed. In this case, the evolution operator generated by $H(t)$ can be shown as
\begin{equation}
	U(t)=U_A(t)= \sum _{n}e^{-i \alpha _{n}(t)+i \gamma _{n}(t)}| \varphi _{n}(t)\rangle \langle \varphi _{n}(0)|,
\end{equation}
where $\alpha_{n}(t)$ is the dynamical phase and $\gamma_{n}(t)$ corresponds to the geometric phase. When the evolution path is cyclic, the geometric phase becomes relevant. That is the main reason why we keep $\gamma_n(t)$ all the way, rather than use the gauge in which all $\braket{\varphi_n(t)}{\dot{\varphi}_n(t)}=0$.

To show how a single-qubit holonomic gate can be achieved, we consider a three-level $\Lambda$-model Hamiltonian which can be written as
\begin{equation}\label{ht}
	H(t)=\Omega(t)\sin\frac{\theta}{2}e^{i\phi}\ket{e}\bra{0}-\Omega(t)\cos\frac{\theta}{2}\ket{e}\bra{1}+\mathrm{H.c.},
\end{equation}
where $\Omega(t)$ is the control envelop while $\theta$ and $\phi$ are time-dependent control parameters (see Fig.~\ref{fig:levels} (a)). We note that in the formula of the Hamiltonians, the parameters $\theta$ and $\phi$ denote $\theta(t)$ and $\phi(t)$ for simplicity while in the formula of the evolution operators we use $\theta$ and $\phi$ to denote $\theta(0)$ and $\phi(0)$. $H(t)$ has three eigenvalues: $E_\pm=\pm\Omega(t)$, and $E_d=0$. The corresponding eigenstates are
\begin{align}
	&\ket{\varphi_+(t)}=\frac{1}{\sqrt{2}}\left(\sin\frac{\theta}{2}e^{-i\phi}\ket{0}-\cos\frac{\theta}{2}\ket{1}+\ket{e} \right),\notag\\
    &\ket{\varphi_-(t)}=\frac{1}{\sqrt{2}}\left(\sin\frac{\theta}{2}e^{-i\phi}\ket{0}-\cos\frac{\theta}{2}\ket{1}-\ket{e} \right),\notag\\
	&\ket{\varphi_d(t)}=\cos\frac{\theta}{2}\ket{0}+\sin\frac{\theta}{2}e^{i\phi}\ket{1}.
\end{align}
Accordingly, we can obtain the corresponding dynamical phases and geometric phases for the three eigenstates
\begin{align}
	&\alpha_{+}= \int _{0}^{t}\Omega(t^{\prime})dt^{\prime}, &
	\gamma_{+}= \int _{0}^{t}\frac{1}{2}\dot{\phi}\sin ^{2}\frac{\theta}{2}dt^{\prime}, \nonumber\\
	&\alpha_{-}=- \int _{0}^{t}\Omega(t^{\prime})dt^{\prime}, &	
	\gamma_{-}= \int _{0}^{t}\frac{1}{2}\dot{\phi}\sin ^{2}\frac{\theta}{2}dt^{\prime}, \nonumber\\
	&\alpha _{d}=0, &
	\gamma_{d}=- \int _{0}^{t}\dot{\phi}\sin ^{2}\frac{\theta}{2}dt^{\prime}.
\end{align}
After a cyclic evolution, the evolution operator $U(t)$ can be shown as
\begin{align}
	U(\tau)=\sum_{k=\pm,d}e^{-i \alpha _{k}(\tau)+i \gamma _{k}(\tau)}\ketbra{\varphi_{k}}{\varphi_{k}},
\end{align}
where $\ket{\varphi_k}=\ket{\varphi_k(0)}$ ($k=\pm,d$).

When $\alpha_{+}(\tau)=-\alpha_{-}(\tau)=\int_{0}^{\tau}\Omega(t)\mathrm{d}t=2k\pi$ ($k$ is an integer) and taking $\gamma_d=-2\gamma_+=-2\gamma_-$ into account, there will be no dynamical phase for the three eigenstates.  Then, the final evolution operator in the basis $\{\ket{e}, \ket{0}, \ket{1}\}$ can be written as
\begin{align}\label{ut}
	U(\tau)=\left(
    \begin{array}{ccc}
    	e^{i\gamma_+}&0&0\\
    	0&\begin{array}{c}
    		e^{i\gamma_+}\sin^2\frac{\theta}{2}\\+e^{-i2\gamma_+}\cos^2\frac{\theta}{2}
    	\end{array}
    	&\begin{array}{c}
    		\frac{1}{2}\sin\theta e^{-i\phi}(e^{-i2\gamma_+}\\-e^{i\gamma_+})
    	\end{array}\\
    	0&
    	\begin{array}{c}
    		\frac{1}{2}\sin\theta e^{i\phi}(e^{-i2\gamma_+}\\-e^{i\gamma_+})
    	\end{array}
    	&\begin{array}{c}
    		e^{i\gamma_+}\cos^2\frac{\theta}{2}\\+e^{-i2\gamma_+}\sin^2\frac{\theta}{2}
    	\end{array}
    \end{array}
	\right),
\end{align}
where $\theta$ and $\phi$ represent $\theta(0)$ and $\phi(0)$, respectively.

It is clear that the final evolution operator $U(\tau)$ can be a proper holonomic quantum gate on the subspace $\mathcal{S}(0)$ spanned
by $\ket{0}$ and $\ket{1}$, as long as the corresponding evolution process satisfies the two holonomic conditions.
First, we can prove that there is no dynamical contribution in the dynamics of $\mathcal{S}(t)$ spanned by $\ket{0(t)}=U(t)\ket{0}$ and $\ket{1(t)}=U(t)\ket{1}$
because
\begin{align}
	\langle j(t)|H(t)|k(t)\rangle =\bra{j}\sum_nE_n(t)\ket{\varphi_n(0)}\braket{\varphi_n(0)}{k} =0,
\end{align}
where $j,k=0,1$.
Second, according to Eq.~(\ref{ut}), if the initial state of the system is in the initial subspace $\mathcal{S}(0)$, it will return to the subspace $\mathcal{S}(0)$ at time $\tau$ since $\mathcal{S}(\tau)=\mathcal{S}(0)$. Therefore, the cyclic evolution driven by $H(t)$ in
Eq.~(\ref{ht}) is pure geometric and the generated $U(\tau)$ is related to a holonomic gate on the subspace $\mathcal{S}(0)$, reading
\begin{align}\label{uh1}
	U_{\mathrm{h1}}=\left(
    \begin{array}{cc}
    	\begin{array}{c}
    		e^{i\gamma_+}\sin^2\frac{\theta}{2}\\+e^{-i2\gamma_+}\cos^2\frac{\theta}{2}
    	\end{array}&
    	\begin{array}{c}
    		\frac{1}{2}\sin\theta e^{-i\phi}(e^{-i2\gamma_+}\\-e^{i\gamma_+})
    	\end{array}\\
    	\begin{array}{c}
    		\frac{1}{2}\sin\theta e^{i\phi}(e^{-i2\gamma_+}\\-e^{i\gamma_+})
    	\end{array}&
    	\begin{array}{c}
    		e^{i\gamma_+}\cos^2\frac{\theta}{2}\\+e^{-i2\gamma_+}\sin^2\frac{\theta}{2}
    	\end{array}
    \end{array}
	\right),
\end{align}
where $\theta$ and $\phi$ represent $\theta(0)$ and $\phi(0)$, respectively.

Compared with the original nonadiabatic holonomic gates which depends only on the initial parameters, $U_{\mathrm{h1}}$ depends on both the initial parameters ($\theta(0)$ and $\varphi(0)$) and the path that the parameters travel in the parameter space, indicating that $U_{\mathrm{h1}}$ is universal for single-qubit gates.
However, the holonomic gates we proposed here offer an extra variable. This implies that an arbitrary single-qubit gate can be built with greater feasibility. Moreover, the path provides an additional degree of freedom for optimal design of holonomic gates that are robust against control errors.

\begin{figure}
	\centering
	\includegraphics[width=0.48\textwidth]{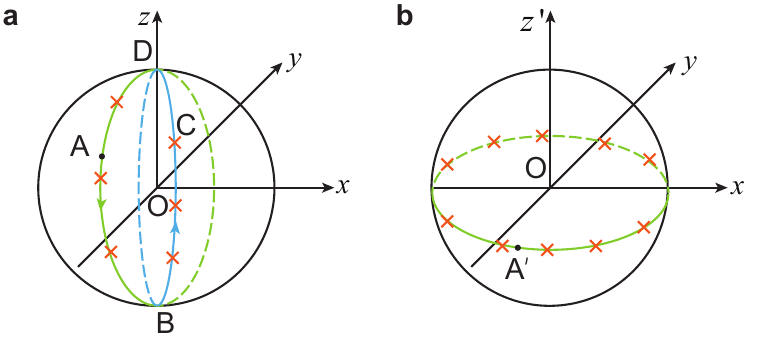}\\
	\caption{\textbf{Evolution path to achieve an arbitrary single-qubit holonomic gate.} The red crosses on the evolution path indicate the points where the $\pi$-pulse based Hamiltonian can be applied. (a) The parameter space is defined by the polar angle $\theta$ and the azimuthal angle $\phi$. The evolution path starts at point ``A'' which is located at $\theta(0)$ and $\phi(0)$. By increasing $\theta(0)$ to $\pi$ while keeping $\phi(0)$ unchanged, the Hamiltonian travels to the South Pole ``B'' along a geodesic line. After changing $\phi(0)$ to the desired $\phi(t)$ (see (b) for details), the Hamiltonian moves to the North Pole ``D'' along another geodesic line. At point ``D'', $\phi(t)$ is then returned to $\phi(0)$, and the Hamiltonian returns to ``A'' along the original geodesic line. The path on the sphere consists of three segments: AB-BCD-DC, each of which is a geodesic line. (b) At point ``B'', the parameter space is defined by a new polar angle $\theta'$ and the same azimuthal angle $\phi$. The vector OA$'$ here has the same azimuthal angle as that for the  vector OA in (a). Changing $\phi(0)$ to $\phi(t)$ requires moving the Hamiltonian from A$'$ to the corresponding point of $\phi(t)$ on the equator.}\label{fig:path}
\end{figure}

Explicitly, when $\gamma_+ = \pi$, $U_{\mathrm{h1}}$ becomes the holonomic gate as obtained in the original time-independent $\Lambda$-model scheme \cite{sjoqvist2012non},
reading
\begin{align}
	U_{\mathrm{h1(xy)}}=\left(
    \begin{array}{cc}
    	\cos\theta & \sin\theta e^{-i\phi}\\
    	\sin\theta e^{i\phi}& -\cos\theta
    \end{array}
	\right).
\end{align}
Notably, the three Pauli operators $\sigma_x$, $\sigma_y$, and $\sigma_z$ correspond to the parameter settings of $\theta(0)=\pi/2$ and $\phi(0)=0$, $\theta(0)=\pi/2$ and $\phi(0)=\pi/2$, and $\theta(0)=0$, respectively.
On the other hand, by choosing $\theta(0)= 0$, we can get a pure geometric rotation around $\sigma_z$ up to a phase factor $e^{-i2\gamma_+}$
\begin{align}
	U_{\mathrm{h1(z)}}=e^{-i2\gamma_+}\left(
     \begin{array}{cc}
     	1& 0\\
     	0& e^{i3\gamma_+}
     \end{array}
	\right),
\end{align}
which can not be obtained directly from the original scheme \cite{sjoqvist2012non}. A summary of the single-qubit gates and the corresponding parameters are provided in Tab.~\ref{tab:parameter}.
\begin{table}[htbp]
	\centering
	\caption{Parameters for the holonomic gates.}
	\label{tab:parameter}
	\begin{tabular}{lc}
		\hline\hline
		Target gate & Parameter setting \\
		\hline
		$\sigma_x$ & $\theta(0)=\pi/2$, $\phi(0)=0$, $\gamma_+=\pi$ \\
		$\sigma_y$ & $\theta(0)=\pi/2$, $\phi(0)=\pi/2$, $\gamma_+=\pi$ \\
		$\sigma_z$ & $\theta(0)=0$, $\phi(0)=0$, $\gamma_+=\pi$ \\
		H          & $\theta(0)=\pi/4$, $\phi(0)=0$, $\gamma_+=\pi$ \\
		S          & $\theta(0)=0$, $\phi(0)=0$, $\gamma_+=\pi/6$ \\
		CNOT       & $\theta(0)=\pi/2$, $\phi(0)=0$, $\gamma_+=\pi$ \\
		Controlled-phase & $\theta(0)=0$, $\phi(0)=0$, $\gamma_+=\gamma/3$ \\
		\hline\hline
	\end{tabular}
\end{table}

\begin{figure*}
	\centering
	\includegraphics[width=1.0\textwidth]{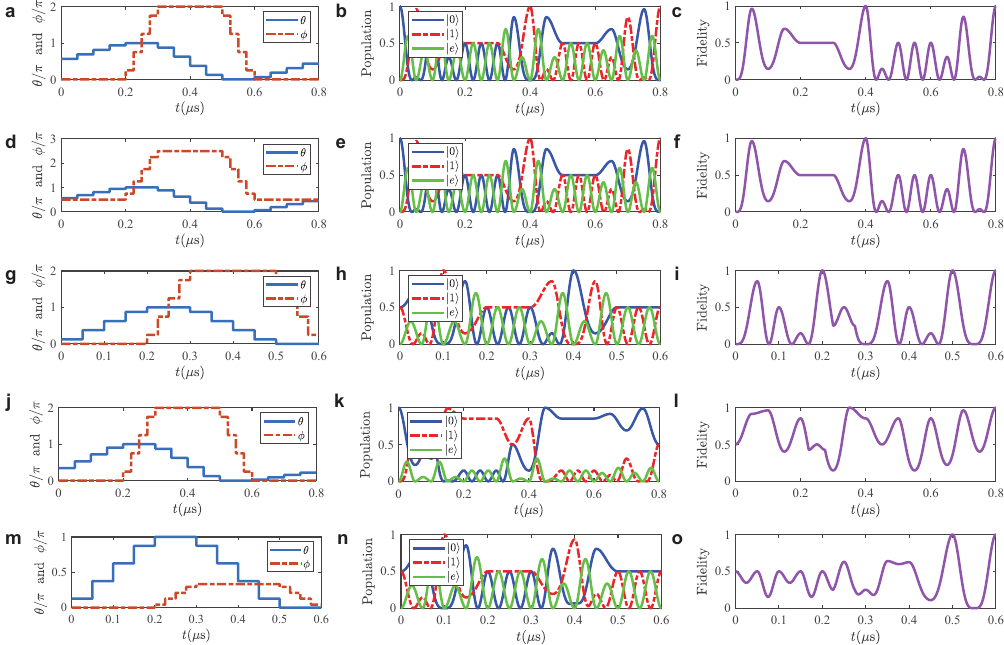}\\
	\caption{\textbf{Single-qubit holonomic gates implemented using the $\pi$-pulse scheme.} From the first to the fifth row, the data correspond to the $\sigma_x$, $\sigma_y$, $\sigma_z$, Hadamard (H), and phase (S) gates, respectively. The initial state is set to $\ket{0}$ for the $\sigma_x$, $\sigma_y$, and H gates, and to $(\ket{0} + \ket{1})/\sqrt{2}$ for the $\sigma_z$ and S gates. The first column illustrates the evolution of the control parameters $\theta$ and $\phi$; the second column shows the population dynamics of the three-level system; and the third column presents the fidelity between the evolving state and the ideal target state.}
	\label{fig:singlegates}
\end{figure*}
\subsection{Evolution path along geodesic lines}

We now detail how to implement the above adiabatic process using geodesic lines in the parameter space. The $\Lambda$-model Hamiltonian $H(t)$ in Eq.~(\ref{ht}) is parameterized by the angles $\theta$ and $\phi$, which define a spherical coordinate system where $\theta$ and $\phi$ correspond to the polar and azimuthal angles, respectively, as illustrated in Fig.~\ref{fig:path}(a). To realize the holonomic gate $U_{h1}$ described in Eq.~(\ref{uh1}), we first specify the parameters $\theta(0)$, $\phi(0)$, and the geometric phase $\gamma_+$. The initial values $\theta(0)$ and $\phi(0)$ determine the starting point ``A'' on the surface of the sphere, while $\gamma_+$ is accumulated along a closed three-segment path ``AB-BCD-DA'', where each segment corresponds to a geodesic curve.

The full control sequence consists of five steps:

\textbf{Step 1:} From point ``A'' to the South Pole ``B'', the polar angle $\theta$ increases from $\theta(0)$ to its maximum value $\pi$, with the azimuthal angle held fixed at $\phi(t)=\phi(0)$.

\textbf{Step 2:} At point ``B'', the azimuthal angle $\phi$ is varied from $\phi(0)$ to a new value $\phi(t)$, such that $\phi(t)-\phi(0)=2\gamma_+$, thereby realizing the desired geometric phase. During this step, the Hamiltonian reduces to an effective two-level system:
\begin{equation}
	H(t)=\Omega(t)e^{i\phi}\ket{e}\bra{0}+\text{H.c.},
\end{equation}
which can be equivalently expressed as
\begin{equation}
	H(t)=\Omega(t)(\cos\phi\,\sigma_x + \sin\phi\,\sigma_y),
\end{equation}
where $\sigma_x = \ketbra{e}{0} + \ketbra{0}{e}$, $\sigma_y = i(\ketbra{e}{0} - \ketbra{0}{e})$, and $\sigma_z = \ketbra{0}{0} - \ketbra{e}{e}$.

To visualize the trajectory during this step, we introduce an auxiliary parameter space, as shown in Fig.~\ref{fig:path}(b). Here, the azimuthal angle $\phi$ remains consistent with the original space, while the effective polar angle $\theta'$ is fixed at $\pi$/2, since the Hamiltonian has no $\sigma_z$ component. This implies that the state evolution proceeds along the equator of the corresponding Bloch sphere.

\textbf{Step 3:} The system then evolves from ``B'' through ``C'' to ``D'' as the polar angle $\theta$ decreases from $\pi$ to $0$, with $\phi(t)$ held constant.

\textbf{Step 4:} At the North Pole ``D'', the azimuthal angle is reset from $\phi(t)$ to its initial value $\phi(0)$. At this point, the Hamiltonian becomes:
\begin{equation}
	H(t) = -\Omega(t)(\ketbra{e}{1} + \ketbra{1}{e}),
\end{equation}
which is independent of $\phi$, allowing the angle to be reset arbitrarily fast. Since $\theta=0$ during this process, the geometric phases $\gamma_k$ ($k=\pm,d$) vanish, and the evolution remains trivial, provided that the associated dynamical phases $\alpha_{\pm}$ are also zero.

\textbf{Step 5:} Finally, the system is driven from point ``D'' back to the initial point ``A'' by increasing $\theta$ from $0$ to $\theta(0)$, completing the closed evolution path.

\subsection{Accelerating adiabatic evolution via the $\pi$-pulse scheme}

Since in each of the five steps from \textbf{Step 1} to \textbf{Step 5}, the system evolves along a geodesic path, the entire adiabatic control process can be accelerated using the $\pi$-pulse scheme. These five geodesic segments can be grouped into two categories: the first includes the segments in \textbf{Step 1}, \textbf{Step 3}, and \textbf{Step 5}, which lie on the spherical surface shown in Fig.~\ref{fig:path}(a); the second includes those in \textbf{Step 2} and \textbf{Step 4}, which lie along the equator of the sphere illustrated in Fig.~\ref{fig:path}(b).

We first consider accelerating the evolution for the first category. In these steps, the system Hamiltonian $H(t)$ yields the following nonadiabatic transition terms $G_{kj}$:
\begin{align}
	G_{+-} &= G_{-+} = -\frac{\dot{\phi}}{2} \sin^2\frac{\theta}{2}, \nonumber \\
	G_{+d} &= G_{-d} = G_{d+}^* = G_{d-}^* \nonumber \\
	&= \frac{e^{i\phi}}{2\sqrt{2}} \left(i\dot{\theta} - \dot{\phi}\sin\theta \right).
\end{align}

For these geodesic paths, the angle $\theta$ varies from $\theta_i$ to $\theta_f$ while $\phi$ remains fixed. Each path is divided into $N$ (an integer) equal segments, and the Hamiltonian is applied only at the midpoint of each segment to flip the projector $P_{kj}$ between $1$ and $-1$. This can be achieved by applying a $\pi$-pulse to $P_{kj}$ at each midpoint where $H(\theta)$ is applied. Under this scheme, if $\theta$ changes at a constant rate, the corresponding nonadiabatic transition Hamiltonian $H_T(\theta)$ remains time-independent between consecutive flips, enabling an exact evaluation of the corresponding transition operator $U_T$.
Moreover, the application of each $\pi$-pulse flips $H_T(\theta)$ between $H_T(\theta_i)$ and $-H_T(\theta_i)$, resulting in the following total evolution operator:
\begin{align}
	U_T(\theta) &= e^{-i\frac{\Delta}{2}H_T(\theta_i)}\cdot e^{i\Delta H_T(\theta_i)}\cdot e^{-i\Delta H_T(\theta_i)} \cdots  \nonumber \\
	& \cdot e^{i(-1)^{N+1}\frac{\Delta}{2} H_T(\theta_i)}= I,
\end{align}
where $\Delta = (\theta_f - \theta_i)/N$.

This result demonstrates that the total nonadiabatic transition is completely canceled by the $\pi$-pulse sequence. The maximum deviation from perfect adiabaticity is governed by the norm $||e^{i\frac{\Delta}{2} H_T(\theta_i)}||$, which can be made arbitrarily small by choosing a sufficiently large $N$.

A similar strategy can be applied to the second category of geodesic paths, where $\theta$ remains fixed and $\phi$ varies from $\phi_i$ to $\phi_f$. Again, the path is divided into $N$ equal segments, and the Hamiltonian is applied at the midpoint of each to flip the relevant $P_{kj}$ between 1 and $-1$. When $\theta = \pi$, the only non-zero nonadiabatic couplings are
\begin{equation}
	G_{+-} = G_{-+} = -\frac{\dot{\phi}}{2}.
\end{equation}
If $\phi$ is varied at a constant rate, $G_{+-}$ and $G_{-+}$ become constants. Thus, the nonadiabatic Hamiltonian $H_T(\phi)$ flips between $H_T(\phi_i)$ and $-H_T(\phi_i)$ through the application of $H(\phi)$. Consequently, the total evolution operator $U_T$ at the end of the path also equals the identity.

To explicitly demonstrate the implementation of single-qubit holonomic gates using the $\pi$-pulse scheme, we adopt experimentally feasible parameters for the Hamiltonian $H(t)$ in Eq.~(\ref{ht}), where the Rabi frequency $\Omega$ is set to $2\pi\times 10$ MHz with a rectangular pulse shape. Five single-qubit gates are simulated using the five-step control protocol illustrated in Fig.~\ref{fig:path}, including the three Pauli operators, the Hadamard (H) gate, and the phase (S) gate. In each step, the corresponding geodesic path is divided into five segments, which already provides an effective suppression of nonadiabatic transitions. As shown in the second column of subfigures in Fig.~\ref{fig:singlegates}, the population of the target state approaches unity, while those of the undesired states tend toward zero. Increasing $N$ further reduces the nonadiabatic transitions. The fidelity between the final state and the target state reaches exactly 1 at the end of the evolution, regardless of the evolution time. In our numerical simulations, a representative initial state is used in the construction of each gate; similar performance is observed for other initial states as well.

\subsection{Two-qubit holonomic gate}
\begin{figure*}
	\centering
	\includegraphics[width=1.0\textwidth]{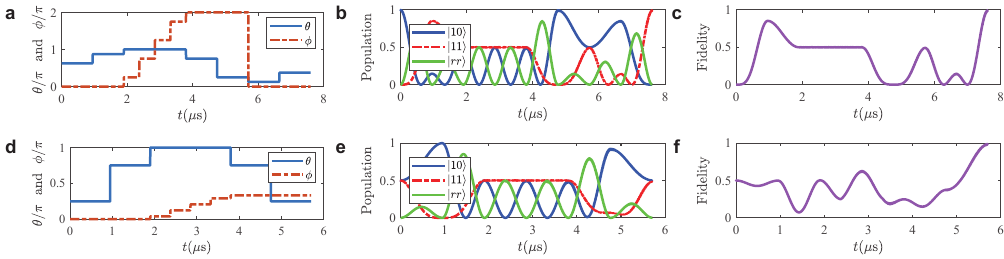}\\
	\caption{\textbf{Two-qubit holonomic gates implemented using the $\pi$-pulse scheme.} The first and second rows correspond to the CNOT gate and the controlled-phase gate, respectively. The initial state for the CNOT gate is $\ket{10}$, while that for the controlled-phase gate is $(\ket{10} + \ket{11})/\sqrt{2}$. The first column illustrates the evolution of the control parameters $\theta$ and $\phi$. The second column shows the population dynamics of the basis states $\ket{10}$, $\ket{11}$, and $\ket{rr}$. The third column presents the fidelity between the evolving state and the ideal target state.}
	\label{fig:twogates}
\end{figure*}
To achieve universal holonomic quantum computation, a nontrivial two-qubit gate is required in addition to single-qubit holonomic gates. To demonstrate the feasibility of our scheme for two-qubit gate implementation, we consider a system composed of two three-level Rydberg atoms as a concrete example \cite{su2017applications,saffman2010quantum}. Each atom has two stable ground states, $\ket{0}$ and $\ket{1}$, and a long-lived Rydberg state $\ket{r}$, as illustrated in Fig.~\ref{fig:levels}(b).

For the first atom, the transition between $\ket{1}$ and $\ket{r}$ is driven by a pair of bichromatic laser pulses with symmetric detunings $-\Delta$ and $+\Delta$, both having the same Rabi frequency $\Omega_{11}(t)$. For the second atom, the Rydberg state $\ket{r}$ is coupled to both ground states $\ket{0}$ and $\ket{1}$ via two separate pairs of bichromatic pulses with detunings $-\Delta$ and $+\Delta$, and with Rabi frequencies $\Omega_{20}(t)$ and $\Omega_{21}(t)$, respectively.

Under the rotating-wave approximation and in the rotating frame, the total Hamiltonian of the two-atom system reads
\begin{align}\label{h2t}
	H(t) = & V_{12} \ket{rr}\bra{rr} + \left[ \Omega_{11}(t)(e^{-i\Delta t} + e^{i\Delta t}) \ket{r}_1 \bra{1} \right. \notag \\
	& + \Omega_{20}(t)(e^{-i\Delta t} + e^{i\Delta t}) \ket{r}_2 \bra{0} \notag \\
	& \left. + \Omega_{21}(t)(e^{-i\Delta t} + e^{i\Delta t}) \ket{r}_2 \bra{1} + \mathrm{H.c.} \right],
\end{align}
where $V_{12}$ denotes the Rydberg-Rydberg interaction strength between the two atoms.

When the condition $\Delta \gg |\Omega_{11}(t)|, |\Omega_{20}(t)|, |\Omega_{21}(t)|$ is satisfied, and the interaction strength is set as
\begin{equation}
	V_{12} = 2\Delta - \frac{4[\Omega_{11}^2(t) + \Omega_{20}^2(t) + \Omega_{21}^2(t)]}{3\Delta},
\end{equation}
we can derive an effective Hamiltonian using the method in Ref.~\cite{james2007effective} (see Appendix~\ref{appb} for details), given by
\begin{align}\label{heff}
	H_{\mathrm{eff}}(t) = \frac{2\Omega_{11}(t)}{\Delta} \left[ \Omega_{20}(t)\ket{rr}\bra{10} + \Omega_{21}(t)\ket{rr}\bra{11} \right] + \mathrm{H.c.}
\end{align}
This effective Hamiltonian, which can also be implemented in NV centers \cite{zu2014experimental} and trapped ions \cite{zhao2019nonadiabatic}, forms a $\Lambda$-type configuration involving the states $\ket{10}$, $\ket{11}$, and the common excited state $\ket{rr}$, analogous to the structure used in the single-qubit case [see Fig.~\ref{fig:levels}(c)]. By setting the effective coupling strengths as
\begin{align}
	\frac{2\Omega_{11}(t)\Omega_{20}(t)}{\Delta} &= \Omega(t)\sin\frac{\theta}{2} e^{i\phi}, \\
	\frac{2\Omega_{11}(t)\Omega_{21}(t)}{\Delta} &= -\Omega(t)\cos\frac{\theta}{2},
\end{align}
the effective Hamiltonian $H_{\mathrm{eff}}(t)$ evolves in the same parameter space $(\theta, \phi)$ as the single-qubit holonomic Hamiltonian in Eq.~(\ref{ht}).

Focusing on the subspace spanned by $\{\ket{10}, \ket{11}, \ket{rr}\}$, a non-Abelian geometric phase can be accumulated during a cyclic evolution in the $(\theta, \phi)$ parameter space. As a result, a holonomic two-qubit gate is realized on the logical subspace $\mathcal{S}'(0) = \text{span}\{\ket{10}, \ket{11}\}$, given by
\begin{equation}
	U_{\mathrm{h2}} = \ket{0}_1\bra{0} \otimes I + \ket{1}_1\bra{1} \otimes U_{\mathrm{h1}},
\end{equation}
where $I = \ket{0}_2\bra{0} + \ket{1}_2\bra{1}$ is the identity operation on the second qubit, and $U_{\mathrm{h1}}$ denotes the single-qubit holonomic gate defined in Eq.~(\ref{uh1}).

Since an arbitrary single-qubit gate can be realized with $U_{\mathrm{h1}}$ in one loop, an arbitrary controlled gate can be obtained with $U_{\mathrm{h2}}$ in one loop too.
For example, by choosing $\gamma_+ = \pi$, $\theta(0) = \pi/2$, and $\phi(0)= 0$, the resulting gate becomes a holonomic controlled-NOT (CNOT) gate between the two Rydberg atoms. Alternatively, when $\theta(0) = 0$, the effective Hamiltonian generates a controlled-phase gate with $3\gamma_+$ as the conditional phase shift.
A summary of the two-qubit gates and the corresponding parameters are provided in Tab.~\ref{tab:parameter}.

To demonstrate the performance of the two-qubit holonomic gates, we numerically simulate the CNOT gate and the controlled-phase gate using experimentally feasible parameters for Rydberg atoms. Specifically, the Rabi frequency $\Omega_{11}$ is set to $2\pi\times10$ MHz, with $\Delta = 38\Omega_{11}$, $\Omega_{20} = \Omega_{11} \sin(\theta/2) e^{i\phi}$, and $\Omega_{21} = -\Omega_{11} \cos(\theta/2)$. As shown in the first column of subfigures in Fig.~\ref{fig:twogates}, the geodesic paths in \textbf{Step 1}, \textbf{3}, and \textbf{5} are discretized into three segments, while that in \textbf{Step 2} is divided into five segments. Here \textbf{Step 4} is omitted to shorten the evolution time and thereby improve the fidelity. Similar to the single-qubit case, nonadiabatic transitions in the evolution of the two-qubit holonomic gates are effectively canceled, and the final gate fidelity approaches unity.

\section{Conclusion}\label{conclusion}
We have introduced a novel scheme for implementing fast and robust holonomic quantum gates by combining adiabatic evolution with the $\pi$-pulse technique. Unlike traditional shortcut-to-adiabaticity or nonadiabatic HQC approaches, our method maintains the adiabatic following of instantaneous eigenstates while significantly accelerating gate operation. The key idea lies in steering the system Hamiltonian along geodesic paths and applying control pulses that suppress nonadiabatic transitions via phase modulation. This approach enables the realization of a universal set of quantum gates, including arbitrary single-qubit and controlled two-qubit gates, within a single evolution loop. Our scheme thus bridges the gap between speed and robustness in HQC and enhances the feasibility of scalable fault-tolerant quantum computing.

In our work, the $\pi$-pulse scheme \cite{wang2016necessary} is employed to accelerate adiabatic evolutions. Recently, this scheme has been experimentally verified \cite{xu2019breaking,zheng2022accelerated,gong2023accelerated} and applied to quantum sensing \cite{zeng2024wide}, electron-nuclear resonance enhancement \cite{xu2024enhancing}, and state preparation \cite{chen2024fast}. Leveraging newly developed $\pi$-pulse schemes to construct holonomic gates represents a promising direction for future research.

\begin{acknowledgments}
	J.Z. acknowledges support by the National Natural Science Foundation of China (No. 12004206).
	T.X. acknowledges support by the National Natural Science Foundation of China (No. 12247165).
	G.L. acknowledges support by the National Natural Science Foundation of China (No. 62471046), Beijing Advanced Innovation Center for Future Chip (ICFC), and Tsinghua University Initiative Scientific Research Program.
\end{acknowledgments}

\appendix
\setcounter{equation}{0}
\section{}\label{appa}
In order to prove $\lim_{\tau\rightarrow +\infty}||F||=0$, we only have to show that
\begin{align}
	F_{kj}(s)= \int_{0}^{s}H_{T,kj}(s^{\prime})\mathrm{d}s^{\prime}\rightarrow 0, \tau\rightarrow+\infty.
\end{align}
(i) when there is no energy crossing, i.e., $E_k(s)\neq E_j(s)$.

\begin{widetext}
	\begin{align}
		F_{kj}(s)=&\int_0^s H_{T,kj}(s^\prime)\mathrm{d}s^\prime \nonumber \\
		=& i\int_0^se^{i\tau[\alpha_k(s^\prime)-\alpha_j(s^\prime)]+i[\gamma_j(s^\prime)-\gamma_k(s^\prime)]}\braket{\dot{\varphi}_k(s^\prime)}{\varphi_j(s^\prime)}\mathrm{d}s^\prime \nonumber \\
		=& \int_0^s\frac{\braket{\dot{\varphi}_k(s^\prime)}{\varphi_j(s^\prime)}}{\tau[E_k(s^\prime)-E_j(s^\prime)]+[i\braket{\varphi_j(s^\prime)}{\dot{\varphi}_j(s^\prime)}-i\braket{\varphi_k(s^\prime)}{\dot{\varphi}_k(s^\prime)}]}
		\mathrm{d}(e^{i\tau[\alpha_k(s)-\alpha_j(s)]+i[\gamma_j(s)-\gamma_k(s)]}) \nonumber \\
		=& \left.\frac{\braket{\dot{\varphi}_k(s^\prime)}{\varphi_j(s^\prime)}}{\tau[E_k(s^\prime)-E_j(s^\prime)]+[i\braket{\varphi_j(s^\prime)}{\dot{\varphi}_j(s^\prime)}-i\braket{\varphi_k(s^\prime)}{\dot{\varphi}_k(s^\prime)}]}
		e^{i\tau[\alpha_k(s^\prime)-\alpha_j(s^\prime)]+i[\gamma_j(s)-\gamma_k(s)]}\right|_0^s \nonumber \\
		&-\frac{1}{\tau}\int_0^se^{i\tau[\alpha_k(s^\prime)-\alpha_j(s^\prime)]+i[\gamma_j(s^\prime)-\gamma_k(s^\prime)]}\frac{\mathrm{d}}{\mathrm{d}s^\prime}(\frac{\braket{\dot{\varphi}_k(s^\prime)}{\varphi_j(s^\prime)}}
		{[E_k(s^\prime)-E_j(s^\prime)]+\frac{1}{\tau}[i\braket{\varphi_j(s^\prime)}{\dot{\varphi}_j(s^\prime)}-i\braket{\varphi_k(s^\prime)}{\dot{\varphi}_k(s^\prime)}]}).
	\end{align}
\end{widetext}

Note that $\braket{\varphi_n(s^\prime)}{\dot{\varphi}_n(s^\prime)}$ are controllable (depending on the evolution path that we design) and limited, thus when $\tau$ goes to infinity, $(\braket{\varphi_j(s^\prime)}{\dot{\varphi}_j(s^\prime)}-\braket{\varphi_k(s^\prime)}{\dot{\varphi}_k(s^\prime)})/\tau$ goes to zero.

Using the Hilbert-Schmidt norm, we have
\begin{widetext}
	\begin{align}
		&\|F_{kj}\|\leq  \frac{1}{\tau}\left\|\frac{\braket{\dot{\varphi}_k(s^\prime)}{\varphi_j(s^\prime)}}
		{[E_k(s^\prime)-E_j(s^\prime)]+\frac{1}{\tau}[i\braket{\varphi_j(s^\prime)}{\dot{\varphi}_j(s^\prime)}-i\braket{\varphi_k(s^\prime)}{\dot{\varphi}_k(s^\prime)}]}\right\| \nonumber \\
		&+\frac{1}{\tau}\left\|\frac{\braket{\dot{\varphi}_k(0)}{\varphi_j(0)}}
		{[E_k(0)-E_j(0)]+\frac{1}{\tau}[i\braket{\varphi_j(0)}{\dot{\varphi}_j(0)}-i\braket{\varphi_k(0)}{\dot{\varphi}_k(0)}]}\right\| \nonumber \\
		&+\frac{1}{\tau}\int_0^s\left\|\frac{\mathrm{d}}{\mathrm{d}s^\prime}(\frac{\braket{\dot{\varphi}_k(s^\prime)}{\varphi_j(s^\prime)}}
		{[E_k(s^\prime)-E_j(s^\prime)]+\frac{1}{\tau}[i\braket{\varphi_j(s^\prime)}{\dot{\varphi}_j(s^\prime)}-i\braket{\varphi_k(s^\prime)}{\dot{\varphi}_k(s^\prime)}]}) \right\|.
	\end{align}
\end{widetext}

Since $E_k(s^\prime)-E_j(s^\prime)$ and $\braket{\varphi_j(s^\prime)}{\dot{\varphi}_j(s^\prime)}$ are independent of total evolution time $\tau$, we have $\|F_{ij}\|=0$ when $\tau$ goes to infinity,. \\~\\

(ii) when there are energy crossings but with a finite number, i.e., eigenvalues $E_k(s)$ and $E_j(s)$ cross at some points.

Without loss of generality, we consider only one cross point (at $s=s_0$) in the first place.
As we discussed in case (i), we can assume that $\|H_{T,kj}\|\leq M$, where $M$ is a constant independent of $\tau$.
In this case, we can choose an arbitrary small $\varepsilon$, and let $\delta=\varepsilon/2M$.
As a result, we can show that
\begin{align}
	F_{kj}=&\int_0^{s_0-\delta}H_{T,kj}(s^\prime)\mathrm{d}s^\prime+\int_{s_0-\delta}^{s_0+\delta}H_{T,kj}(s^\prime)\mathrm{d}s^\prime \nonumber \\
	&+\int_{s_0+\delta}^{s}H_{T,kj}(s^\prime)\mathrm{d}s^\prime=F_\delta^-+F_\delta+F_\delta^+
\end{align}
From case (i), we have
\begin{equation}
	\lim_{\tau\rightarrow+\infty}F_\delta^-=\lim_{\tau\rightarrow+\infty}F_\delta^+=0.
\end{equation}
On the other hand, it is clear that
\begin{equation}
	\|F_\delta\|\leq\varepsilon,
\end{equation}
which can be arbitrary small.
Adding the three terms together, we have
\begin{equation}
	\lim_{\tau\rightarrow+\infty}F=0.
\end{equation}
The above proof can be directly generalized to the case where the number of energy crossings are finite.

\section{}\label{appb}

In this Appendix, we show how to derive the effective Hamiltonian $H_{\mathrm{eff}}(t)$ in Eq.~(\ref{heff}) from $H(t)$ in Eq.~(\ref{h2t}).

The Hamiltonian shown in Eq.~(\ref{h2t}) can be rewritten as
\begin{align}
	&H(t)=\Omega_{11}(t)(e^{-i\Delta t}+e^{i\Delta t})(\ket{r0}\bra{10}+\ket{r1}\bra{11}+\ket{rr}\bra{1r})\notag\\
	&+\Omega_{20}(t)(e^{-i\Delta t}+e^{i\Delta t})(\ket{0r}\bra{00}+\ket{1r}\bra{10}+\ket{rr}\bra{r0})\notag\\
	&+\Omega_{21}(t)(e^{-i\Delta t}+e^{i\Delta t})(\ket{0r}\bra{01}+\ket{1r}\bra{11}+\ket{rr}\bra{r1})\notag\\
	&+\mathrm{H.c.}+V_{12}\ket{rr}\bra{rr},
\end{align}
For convenience, one can move to another picture with the transform operator $U_{r}=\exp(-i2\Delta t\ket{rr}\bra{rr})$, then Eq.~(\ref{h2t}) can be written as
\begin{align}
	H(t)=
	& \Omega_{11}(t)(e^{-i\Delta t}+e^{i\Delta t})(\ket{r0}\bra{10}+\ket{r1}\bra{11})\notag\\
	&+\Omega_{11}(t)(e^{i\Delta t}+e^{i3\Delta t})\ket{rr}\bra{1r}\notag\\
	&+\Omega_{20}(t)(e^{-i\Delta t}+e^{i\Delta t})(\ket{0r}\bra{00}+\ket{1r}\bra{10})\notag\\
	&+\Omega_{20}(t)(e^{i\Delta t}+e^{i3\Delta t})\ket{rr}\bra{r0}\notag\\
	&+\Omega_{21}(t)(e^{-i\Delta t}+e^{i\Delta t})(\ket{0r}\bra{01}+\ket{1r}\bra{11})\notag\\
	&+\Omega_{21}(t)(e^{i\Delta t}+e^{i3\Delta t})\ket{rr}\bra{r1}+\mathrm{H.c.}\notag\\
	&+(V_{12}-2\Delta)\ket{rr}\bra{rr}\notag\\
   =&\big{\{}\big{[}\Omega_{11}(t)(\ket{r0}\bra{10}+\ket{r1}\bra{11})\notag\\
    &+\Omega_{20}(t)(\ket{0r}\bra{00}+\ket{1r}\bra{10})\notag\\
    &+\Omega_{21}(t)(\ket{0r}\bra{01}+\ket{1r}\bra{11})+\mathrm{H.c.}\big{]}\notag\\
    &+\Omega_{11}(t)^{*}\ket{1r}\bra{rr}\notag\\
    &+\Omega_{20}(t)^{*}\ket{r0}\bra{rr}
    +\Omega_{21}(t)^{*}\ket{r1}\bra{rr}\big{\}}e^{-i\Delta t}\notag\\
    &+\big{[}\Omega_{11}(t)^{*}\ket{1r}\bra{rr}
    +\Omega_{20}(t)^{*}\ket{r0}\bra{rr}\notag\\
    &+\Omega_{21}(t)^{*}\ket{r1}\bra{rr}\big{]}e^{-i3\Delta t}+\mathrm{H.c.}\notag\\
    &+(V_{12}-2\Delta)\ket{rr}\bra{rr}\notag\\
   =&\sum_{n=1}^{2}\hat{h}_ne^{-i\omega_n t}+\hat{h}^\dagger_ne^{i\omega_n t}+(V_{12}-2\Delta)\ket{rr}\bra{rr},
\end{align}
where
\begin{align}\label{hath1}
	\hat{h}_1=
    &\big{[}\Omega_{11}(t)(\ket{r0}\bra{10}+\ket{r1}\bra{11})\notag\\
    &+\Omega_{20}(t)(\ket{0r}\bra{00}+\ket{1r}\bra{10})\notag\\
    &+\Omega_{21}(t)(\ket{0r}\bra{01}+\ket{1r}\bra{11})+\mathrm{H.c.}\big{]}\notag\\
    &+\Omega_{11}(t)^{*}\ket{1r}\bra{rr}\notag\\
    &+\Omega_{20}(t)^{*}\ket{r0}\bra{rr}
    +\Omega_{21}(t)^{*}\ket{r1}\bra{rr},
\end{align}
\begin{align}\label{hath2}
	\hat{h}_2=\Omega_{11}(t)^{*}\ket{1r}\bra{rr}
    +\Omega_{20}(t)^{*}\ket{r0}\bra{rr}
    +\Omega_{21}(t)^{*}\ket{r1}\bra{rr},
\end{align}
and
\begin{align}\label{omegan}
\omega_1=\Delta, \omega_2=3\Delta,
\end{align}

In Ref. \cite{james2007effective}, for a Hamiltonian of the form
\begin{equation}
	H(t)=\sum_{n=1}^{N}\hat{h}_ne^{-i\omega_n t}+\hat{h}^\dagger_ne^{i\omega_n t},
\end{equation}
the corresponding effective Hamiltonian can be written as
\begin{equation}\label{heff1}
	H_{\mathrm{eff}}(t)=\sum_{m,n=1}^{N}\frac{1}{2}\left(\frac{1}{\omega_m}+\frac{1}{\omega_n}\right)\left[\hat{h}^\dagger_m,\hat{h}_n\right]e^{i(\omega_m-\omega_n)t}
\end{equation}
when the condition $\Delta\gg|\Omega_{11}(t)|,|\Omega_{20}(t)|,|\Omega_{21}(t)|$ is satisfied.

By substituting Eqs.~(\ref{hath1})-(\ref{omegan}) into Eq.~(\ref{heff1}), and eliminating the terms that act trivially on the computation space, we have
\begin{align}\label{heffvrr}
	&H_\mathrm{eff}(t)=\frac{2\Omega_{11}(t)\Omega_{20}(t)}{\Delta}\ket{rr}\bra{10}+\frac{2\Omega_{11}(t)\Omega_{21}(t)}{\Delta}\ket{rr}\bra{11}\notag\\
	&+\mathrm{H.c.}\notag\\
	&+(V_{12}-2\Delta+\frac{4[\Omega_{11}(t)^2+\Omega_{20}(t)^2+\Omega_{21}(t)^2]}{3\Delta})\ket{rr}\bra{rr}.
\end{align}
We can set $V_{12}=2\Delta-4[\Omega_{11}(t)^2+\Omega_{20}(t)^2+\Omega_{21}(t)^2]/(3\Delta)$ to obtain the resulting effective Hmiltonian in Eq.~(\ref{heff}).

\end{document}